# Forecasting Models for Daily Natural Gas Consumption Considering Periodic Variations and Demand Segregation

Ergün Yükseltan, Ahmet Yücekaya*, Ayşe Hümeyra Bilge, Esra Ağca Aktunç

Industrial Engineering Department, Kadir Has University, Istanbul, Turkey
*ahmety@khas.edu.tr

**Abstract**

Due to expensive infrastructure and the difficulties in storage, supply conditions of natural gas are different from those of other traditional energy sources like petroleum or coal. To overcome these challenges, supplier countries require take-or-pay agreements for requested natural gas quantities. These contracts have many pre-clauses; if they are not met due to low/high consumption or other external factors, buyers must completely fulfill them. A similar contract is then imposed on distributors and wholesale consumers. It is thus important for all parties to forecast their daily, monthly, and annual natural gas demand to minimize their risk. In this paper, a model consisting of a modulated expansion in Fourier series, supplemented by deviations from comfortable temperatures as a regressor is proposed for the forecast of monthly and weekly consumption over a one-year horizon. This model is supplemented by a day-ahead feedback mechanism for the forecast of daily consumption. The method is applied to the study of natural gas consumption for major residential areas in Turkey, on a yearly, monthly, weekly, and daily basis. It is shown that residential heating dominates winter consumption and masks all other variations. On the other hand, weekend and holiday effects are visible in summer consumption and provide an estimate for residential and industrial use. The advantage of the proposed method is the capability of long term projections and to outperform time series methods.

1. ## Introduction

Natural gas is a widely used energy source for industrial and residential consumption and unlike other energy sources, such as coal or oil; it is transported via expensive pipeline networks that require high capital investment from the producing countries and commitment from the demanding countries. Natural gas projects require significant investments as well as political support and international acceptance. In order to make such an investment feasible, the revenue should be secured through bilateral contracts between suppliers and consumers. In these contracts, an annual purchase commitment is submitted to the exporting country for a relatively long period and a take-or-pay agreement is part of this contract. In such contracts, there are upper and lower bounds for the quantities to be used and the purchaser is obligated to pay for any unused amount. This cost has to be reflected to end-consumers in an effort to balance supply and demand. Therefore, accurate forecasts for their consumption over at least a one-year horizon are of prime importance. On the other hand, the availability of an even limited amount of storage might alleviate the severity of the costs due to failures to forecast accurately.

Natural gas imported or produced by the main supplier and is sold to large industrial plants, industrial parks, and regional distributors, who deliver it to the end users in their defined region. The purchasers are tied by contracts in which they are required to submit a set of hourly, daily, monthly, and annual amounts to be used over the year; they are allowed to withdraw natural gas in a predefined range, and they are also penalized for over- and under-consumption. The main idea is to reach a point where the overall demand is supplied from the national and international pipeline without any penalty that might occur due to the take-or-pay conditions. Therefore, it is crucial to forecast natural gas demand at various time scales, over about a one-year horizon, for a mixture of residential and industrial customers with different consumption profiles. In this paper, we develop a forecasting model consisting of a modulated expansion in Fourier series, supplemented by deviations from comfortable temperatures as a regressor, and apply this model to forecast demand for four major metropolitan areas in Turkey. In all areas, there are both residential and industrial consumers but natural gas is mainly used for heating; hence, it reaches high levels in winter and has high amplitude fluctuations. In winter, the differences in weekday/weekend consumption are masked by fluctuations due to weather conditions, but in summer the weekend/weekday consumption ratios are used to estimate the proportion of industrial consumption, as we discuss in detail later.

The advantages and disadvantages of contracts with take-or-pay provisions for natural gas have been discussed since the 1980s. Masten and Crocker (1985) develop a simple take-or-pay model for commercial agreements between the natural gas producer and the companies that own and operate the pipeline. The model offers a study of how much natural gas should be purchased according to the take-or-pay conditions. Masten (1988) presents the theory of take-or-pay contracts and the rules developed by the state in this regard. Hubbard and Weiner (1986, 1991) develop models for contracts with take-or-pay conditions between natural gas producers and pipeline owners in the United States (US), stating that take-or-pay condition is an assurance for long-term contracts. Schultz (1997) develops a mixed-integer programming model to optimize the economic trade-offs regarding the take-or-pay decisions with and without make-up provisions, that allow the buyer to credit the penalty payments against future "takes" for a limited time interval. Creti and Villeneuve (2004) examine the natural gas contracts for the European Union countries; they indicate the role take-or-pay condition plays in these agreements and discuss how the US will benefit from previous experiences in take-or-pay conditioned contracts.

Research on natural gas demand forecasts is mostly focused on the total consumption of a country. Soldo (2012) presents a review of the methods used for natural gas consumption forecasting. To forecast natural gas demand for Turkey, Ediger and Akar (2007) and Erdogdu (2010) use the ARIMA model, Akkurt et al. (2010) use time series, Melikoglu (2013) uses logistic equations, and Akpinar and Yumusak (2016) use Seasonal Time Series Methods. Li et al. (2011) use a system dynamics model to forecast natural gas demand for China, considering factors such as production strategies, industry policy, GDP growth, infrastructure construction, and changing demand patterns. Fouquet et al. (1997) develop a co-integration model for the natural gas forecast of the United Kingdom (UK) whereas Zhu et al. (2015) apply support vector regression method to the data from the UK. Parikh et al. (2007) develop a model that is based on the GDP growth of India. Gutiérrez et al. (2005) use a stochastic Gompertz innovation diffusion model for Spain. Chen et al. (2018) develop short-term predictions using a Functional Autoregressive model with exogenous variables (FARX) to forecast day-ahead natural gas-flow in Germany. Some of the other methods proposed to forecast natural gas consumption are Bayesian Model Averaging, intelligent Grey model, support vector regression, econometric modelling, and fuzzy-stochastic frontier analysis for countries such as China, Pakistan, Bangladesh, and UAE (Xu and Wang, 2010; Zhang and Yang, 2015; Zeng and Li, 2016; Khan, 2015; Azadeh et al., 2011; Wadud et al., 2011).

There are also studies that aim to forecast sectoral consumption. Mayer and Benjamini (1978), and Liu and Lin (1991) develop models that are based on temperature changes for residential and commercial areas. Sanchez-Ubeda and Berzosa (2007) propose a statistical decomposition-based medium-term forecasting model for the natural gas demand of industrial consumers in Spain and produce daily demand forecasts using a combination of trend, seasonality, and daily variations. Moraes et al. (2008) propose a mathematical formulation for take-or-pay contracts in the Brazilian long-term energy planning that targets a smaller expected operation cost for the national system. Forouzanfar et al. (2010) propose a logistic equation based approach to forecast natural gas consumption for residential and commercial sectors in Iran. Bianco et al. (2014) use linear regression analysis to forecast future consumption of the Italian residential sector. There are also some efforts to forecast regional natural gas consumption. Szoplik (2015) proposes an artificial neural network approach to forecast natural gas consumption in Szczecin, Poland. The model includes calendar and weather effects and presents an analysis of the gas consumption of individual consumers and small industry. Taspinar et al. (2013) use seasonal auto-regressive integrated moving average with additional variables (SARIMAX), artificial neural networks with multilayer perceptron (ANN-MLP) and with Radial Basis Function (ANN-RBF), and multivariate Ordinary Least Squares regression (OLS) to forecast daily and regional natural gas demand for analyzed regions in Turkey. They also consider meteorological data for the short term forecast. Sarak and Satman (2003) use the heating degree-day method to forecast natural gas consumption by residential heating for the cities located nearby natural gas pipelines in Turkey. Durmayaz et al. (2000) use a degree-hour method to estimate the energy requirement for residential heating and fuel consumption in Istanbul whereas Gumrah et al. (2001) use a degree-day approach to model natural gas demand for Ankara using the annual number of customers, average degree-days, and the usage per customer. Demirel et al. (2012) use neural networks and multivariate time series methods to forecast total consumption in Istanbul and analyze the effect of price on consumption.

As opposed to time series methods that are widely used in the literature, such as auto-regressive moving average (ARMA) models, linear models reflect the cause and effect relationship of the variations in the data; hence, they are useful for long-term predictions and planning. Our experience in data analysis demonstrates that it is preferable to model deterministic variations by a linear model and then, to use, if necessary, time series techniques to the residuals. We propose a linear model as a modulated Fourier series expansion to forecast natural gas consumption. We present three options for forecasting:

using (1) only past data, (2) past data with temperature, (3) past data with temperature and feedback, and we show that the forecasting accuracy increases at each step. Residential natural gas demand is highly correlated with heating requirements and fluctuations due to weather conditions dominates and masks all other variations in winter seasons. The modulated Fourier series expansion replicates the seasonal variation and the weekend effects in summer. But, as the temperature comes into play in cold weather, the first forecasting option is supplemented by the deviations from comfortable temperatures as a regressor in the second forecasting option. In order to improve the forecast accuracy further, we propose a day-ahead feedback approach and reach an acceptable forecast accuracy using the third forecasting option. The feedback mechanism is also important for deregulated energy markets in which natural gas can be purchased from the market through auctions. If the natural gas demand can be forecast for the days ahead, a proper demand analysis can be made, and the supply process can be smooth possibly at a reasonably lower cost. We show that our models are satisfactory for the prediction of weekly and monthly demand, which is achieved within 5-10% Mean Absolute Percentage Error (MAPE) and outperforms ARIMA time series model. We also discuss the share of the industrial and residential demand based on the stable demand during summer months, when residential heating use is at its minimum and when there are holidays in summers (from 2006 onwards) during which most of the industrial plants are not operational. This analysis provides a baseline for residential consumption in summer and can be used to estimate the ratio of industrial and residential demand in a city. This information would especially help to plan demand to be declared in the take-or-pay contracts if a change in industrial production is expected in a city.

The proposed models and approaches have the advantage of providing reliable daily, weekly and monthly forecasts over a one-year horizon. The possibility of long term predictions is the main distinctive feature of our methodology which is applicable to any regional or national consumption data. The assessment of demand segregation, which has not been addressed in the literature, is another distinctive feature of the present work. Weekend effects that can be seen only in summer may be used to estimate industrial demand, but, for Turkey, exclusively household consumption levels are the ones during religious holidays, which shift back by 10 days each year. These holidays, that occurred in summer from 2006 onward in the available data, provided a statistically meaningful sample for the estimation of the share of the industrial demand in the aggregate data.

The remainder of this paper is organized as follows. Section 2 provides a description of the data and methodology. The formulation of the model that considers only temperature effect is introduced in Section 3; and of the model that considers both temperature and feedback is presented in Section 4. The discussion on the possibility of segregation of industrial demand is given in Section 5. Finally, in Section 6, the concluding remarks are provided.

## 2. Description of the Data and Methodology

Natural gas consumers have to forecast their daily, monthly, and annual minimum and maximum natural gas withdrawals from the system and submit a contract for the procurement. The contract also shows the total amount of natural gas to be withdrawn from the system. Annual withdrawal is limited by the total of maximum monthly withdrawals. If more natural gas than the maximum withdrawal is demanded, then it is possible to experience a shortage or extra cost. On the other hand, the contract should include minimum withdrawals as the main supplier will plan the amount of natural gas in the system based on the given data.

The most important parameter in such a system is the monthly natural gas demand. The demand depends on both residential use and industrial consumption and possesses stochastic features. Residential demand has a high seasonality effect, and the main determinant is the temperature. Industrial use includes demand for manufacturing facilities, industrial parks, and other small and medium enterprises. As a result, temperature, economic growth, price, and population growth have effects on the demand, and nondeterministic features make the forecasting challenging.

Natural gas contracts in Turkey are examples of take-or-pay agreements. In Turkey, procurement, distribution, tariff determination, and wholesale of natural gas are managed by the main supplier company, Petroleum Pipeline Corporation (BOTAS). The storage of natural gas is limited in Turkey with a total capacity of 2.84 Billion Sm³, which is around 5% of the total annual demand. It is thus important to be able to make predictions within this precision band. In addition to the need to minimize the risks associated with take-or-pay contracts, the liberalization of the natural gas market necessitates short-term predictions on a daily basis. The characteristics of natural gas consumption have a strong dependency on seasonality and work habits. We have analyzed the consumption data of major cities in Turkey in an effort to discover common patterns. The data includes daily consumption values for the period 2002-2017 of Istanbul West, Istanbul East, Ankara, Eskisehir, and Bursa. The natural gas network started to be built in Ankara, in 1988; and in Istanbul, in 1992. By the 2000s, the network infrastructure was more or less stabilized in these two areas, compared to other places. In particular, data for Istanbul West will be used to illustrate the demand forecast, as this part of the city has a well-established infrastructure, a mix of industrial and residential demand, and has a stable demand growth.

In the early 2000s, the distribution of natural gas was very limited and mostly belonged to industrial zones. As natural gas is adopted as the main energy source for heating and daily activities in residential places, an increase is observed in consumption. This increasing trend in demand is observed in the data sets for all the major cities. This trend is due to the expansion of the natural gas network in the early stages and then it is governed by demographics and economic growth in later stages. In Figure 1, we present the data set for Istanbul East, Istanbul West, Ankara, and Eskisehir to give an idea on the progress of natural gas consumption from 2000 onward. One can see that, for Istanbul and Ankara, where the infrastructure started to be built earlier, the consumption has doubled over a 15-year period, while for Eskisehir, it has a five-fold increase, due to the expansion of the natural gas distribution network.

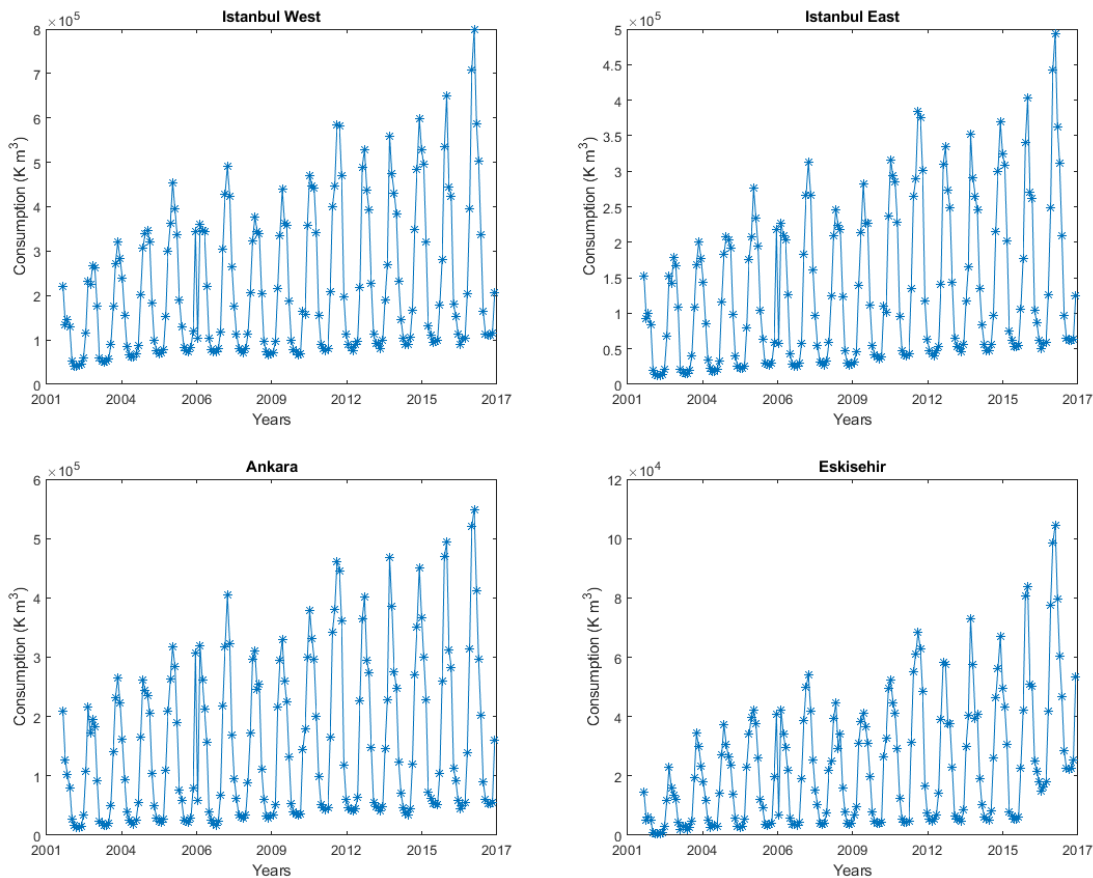

**Figure 1:** Monthly average consumption profiles of four metropolitan areas (2002-2017)

The most remarkable characteristic of the data is the high and irregular consumption in winters as opposed to low and regular consumption in summer. A close-up to the demand data for the year 2016 is given as an example in Figure 2 below. One can see that the consumption in summer is lower and more regular, and displays a clear weekend effect, reflecting the proportion of residential and industrial customers. Winter consumption is higher and much more irregular with large fluctuations, revealing the dominant use of natural gas for heating. Based on this observation, one can clearly state that the daily temperature in winter is the main parameter that affects natural gas consumption. As we consider the daily consumption data, we will use the harmonics of weekly and annual variations as regressors to take into account the weekend effect observed in summer. Yet, as the pattern suggests, the temperature will be the crucial regressor for modelling winter consumption.

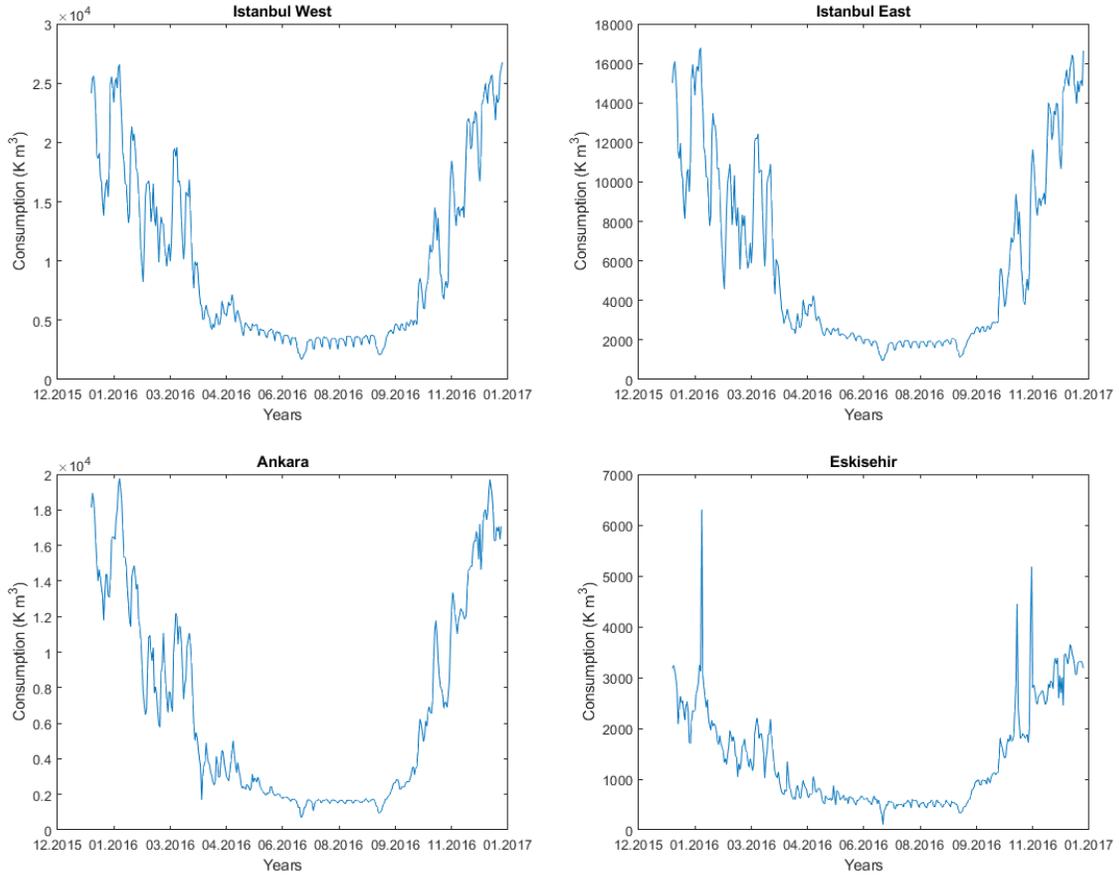

**Figure 2:** One-year consumption profile for four metropolitan areas in 2016

In Figure 2, the low consumption days correspond to religious holidays, which are representative of household-only part of the consumption; and they provide valuable information on demand segregation. Furthermore, especially in Istanbul, the weekend effect is clearly seen in summer.

### 2.1 Fourier Series Expansion and model parameters

The Fourier series expansion of a periodic signal is a well-known tool for mathematical analysis. Under quite mild assumptions, a periodic function *f(t)* with period *T*, can be represented as a sum of sinusoidal functions of periods *T/k, k=1,…,∞*, in the sense that the series converges to *f(t)* pointwise, except at the points of continuity of *f(t)* (Apostol, 1993). The sinusoids whose periods are *T/k* are called the $k^{th}$ harmonics. If the shape of *f(t)* is far from a sinusoidal function, higher harmonics may be stronger than the main variation. It is also well known that if data is sampled at intervals Δ*t*, then the highest harmonic that can be included in the expansion has period *2Δt* (Zayed, 1993). Therefore, periodic variations in observed data, with periods up to *2Δt* can be faithfully represented by a linear model consisting of sine and cosine functions with appropriate periods. Changes in the amplitudes of the periodic variations can be taken into account by multiplying corresponding sinusoids by appropriate functions. For example, the increase in the amplitudes of the seasonal harmonics in time (spanning a couple of years) is taken into account by adding products of seasonal harmonics with a linear term. It is also possible to model periodic changes in the amplitudes of the high frequency variations by adding products of low frequency and high frequency harmonics, but such terms were not included in the present model.

Since data consists of daily averages, the unit time is day; and regressors are represented as column vectors with length equal to the total duration of the observation period in days. The daily gas demand is denoted by S. We use a constant vector (denoted by 1) and a linear term (denoted by *t*) to take into account the trend in the data. Periodic variations consist of $X_n$ and $Z_m$ which are the $n^{th}$ harmonics of sinusoidal functions with a period of one year, i.e., *364/n* days, and the $m^{th}$ harmonics of one week, i.e., *7/m* days, respectively. The regressors that represent the increase in the amplitudes of the periodic variations are the component-wise products of the corresponding vectors by a linear term that we denote as *t*. The Fourier Series Expansion (FSE) model, consisting of a linear trend, seasonal and weekly harmonics and modulation terms can be used to forecast the demand reasonably well over a period encompassing a few years, but it fails to follow irregular variations especially in winter. This is due to the fact that natural gas consumption has a strong dependency on weather conditions. In the present work, we incorporate the effect of climatic conditions into a single parameter, the deviations from comfortable temperatures, defined as follows. For a given temperature *T*, the regressor ($T_d$) is defined as deviations from a comfortable temperature ($T_c$) as

$$T_d = \max(T_c - T, 0) \quad (1)$$

These regressors, and $T_d$ defined above, are arranged as the columns of a matrix *F,* to form the Fourier Series Expansion with Temperature (FSET), as below

$$F = [1 \quad t \quad X_1 \quad X_2 \quad ... \quad X_K \quad Z_1 \quad Z_2 \quad ... \quad Z_M \quad X_1 t \quad ... \quad X_p t \quad T_d] \quad (2)$$

where *p<K*. The number of regressors should be large enough to capture the main features of the data, but over-specification should be avoided. In addition, by the sampling theorem, the shortest allowable period is two days. By taking these considerations into account, we have built a model that uses 30 time regressors (with *K=24* and *M=6*) to represent sinusoidal variations, and 10 regressors (*p=10*) to implement modulation effects. The components of the FSE model are given below.

Polynomial part:

$$Y_t^{(1)} = a + bt \quad (3)$$

Seasonal harmonics:

$$Y_t^{(2)} = \sum_{n=1}^{12} A_n \sin(n\alpha t) + B_n \cos(n\alpha t) \quad (4)$$

where $\alpha = 2\pi/364$ in which we use 364 days for better numerical resolution.

Weekly harmonics:

$$Y_t^{(3)} = \sum_{n=1}^{2} C_n \sin(n\beta t) + D_n \cos(n\beta t) \tag{5}$$

where $\beta = 2\pi/7$

Modulation terms:

$$Y_t^{(4)} = \sum_{n=1}^{5} H_n \sin(n\alpha t)\, t + K_n \cos(n\alpha t)\, t \tag{6}$$

The coefficient vector $a$ and model vector $y$ are calculated as follows.

$$a = (F^t F)^{-1} F^t S \tag{7}$$

$$y = Fa \tag{8}$$

For predicting data, we split the time axis into "past" ($t_1$) and "future" ($t_2$). Once we choose the splitting of the time axis into past and future, we have a corresponding splitting of the matrix $F$, into $F_1$, $F_2$. In order to make a prediction, we use $F_1$ to compute the coefficient vector $a$, but we use $F_2$ to compute the model vector $y_2$ as follows.

$$a_1 = (F_1^t F_1)^{-1} F_1^t S_1 \tag{9}$$

$$y_2 = F_2 a_1 \tag{10}$$

In the assessment of the performance of the models, we use Root Mean Square Error (RMSE) and Mean Absolute Percentage Error (MAPE). If $S_h$ and $R_h$ are the actual demand and the forecast demand for period $h$, respectively, then RMSE and MAPE can be defined as

$$RMSE = \sqrt{\frac{1}{H} \sum_{h=1}^{H} (R_h - S_h)^2} \tag{11}$$

$$MAPE = \frac{100}{H} \sum_{h=1}^{H} \frac{|R_h - S_h|}{S_h} \tag{12}$$

where $H$ is the total number of estimated values. Although we calculate MAPE and RMSE values for monthly, weekly and daily forecasting results, we present the results in terms of percentage for the sake of clarity. We present both FSE and FSET models in an effort to show the effect of temperature on the forecasting accuracy and provide an option to forecast demand using FSE model when the temperature data is not available. We provide the details of each model and accuracy levels in the following sections.

### 3. Forecasting Using Fourier Series Expansion with Temperature

In this section, we present the features of the data on annual, monthly, weekly, and daily timescales, and present the results of models for these variations.

Annual consumption for the four cities over 17 years shows that the growth trend is initially high, and then it decreases. As these are due to a transient phase of network expansion, no modelling is attempted for the whole range. Furthermore, population growth and industrialization are not included in the model. As a typical example, in Figure 3 below, we provide a plot of total consumption, winter consumption, and summer consumption for Istanbul West.

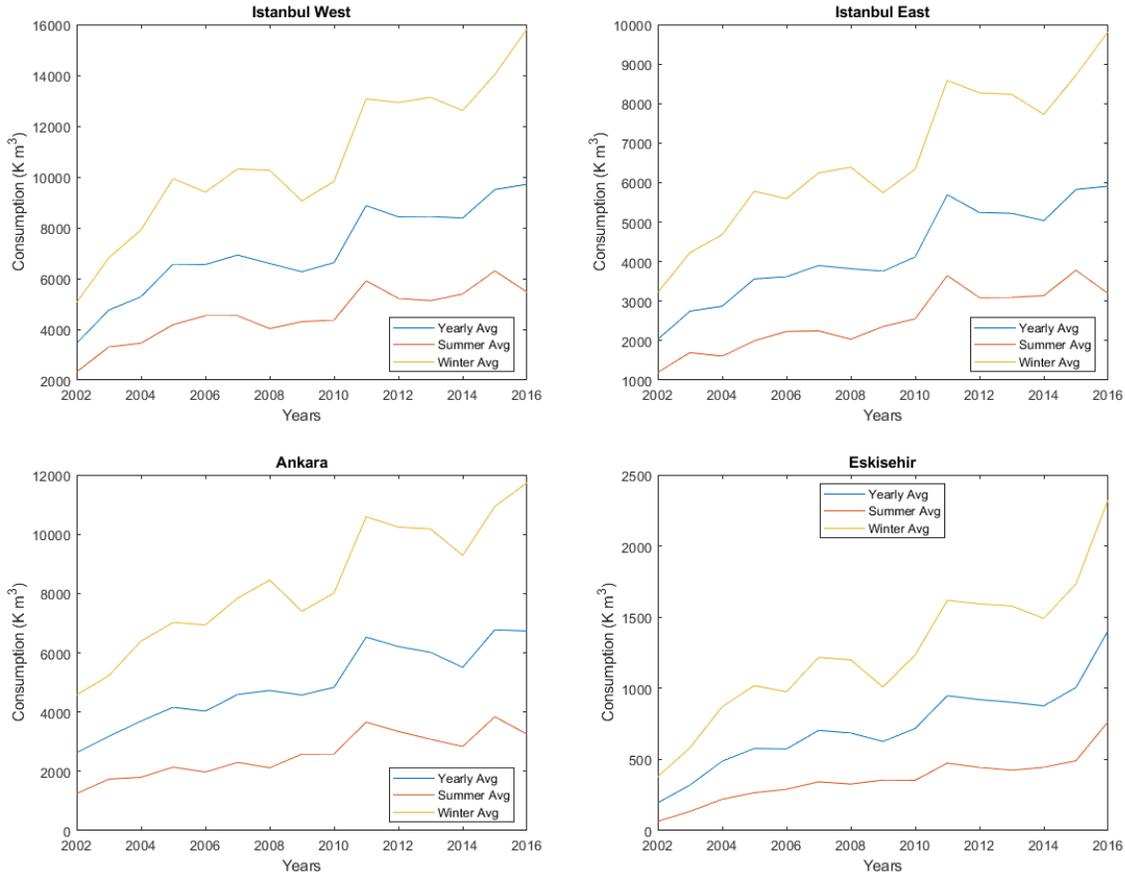

**Figure 3:** Average consumption for four cities; annual mean, mean of summer (months 4-9) and winter (months 1-3 and 10-12)

### 3.1 Modeling of monthly variations

Monthly consumption planning is a complex process that should take into account the expansion of the network and climatic conditions. Summer consumption is more reliable for the determination of trend lines. One should, however, update these trend estimates using the information on the planning of industrial plants.

For prediction purposes, we model monthly variations as a time series. These variations are modeled by using harmonics of one year and monthly mean temperature as regressors based on the temperature data obtained from the Weather Underground website, and the result for Istanbul West is displayed in Figure 4 below. As seen in this figure, consumption profiles are now smoother than daily consumption profiles. For the details of the modeling process, we refer to the section on the modeling of daily variations. Modeling errors for the four cities in monthly demand forecasting are shown in Table 1.

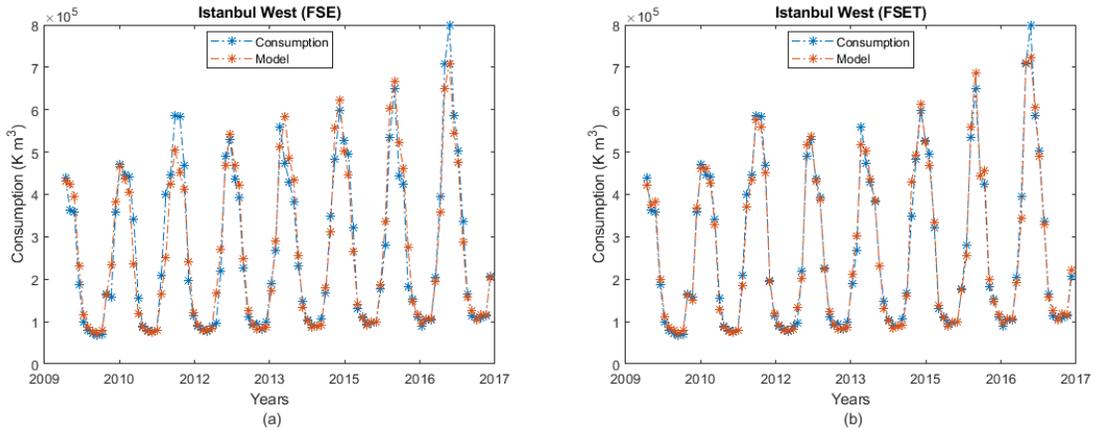

**Figure 4:** Forecasting of monthly consumption for Istanbul West using (a) FSE and (b) FSET

### 3.2 Modeling of Weekly Variations

In this section, data is rearranged on a weekly basis. The forecasting of weekly demand for Istanbul West is given in Figure 5. The holiday periods are clearly seen, and as will be explained later, these periods can be used for demand segregation. The model is adapted to weekly variations, and weekly mean temperature is used as a regressor. It can be seen in Figure 5 that our weekly modeling also follows the pattern well. The results are presented in Table 1.

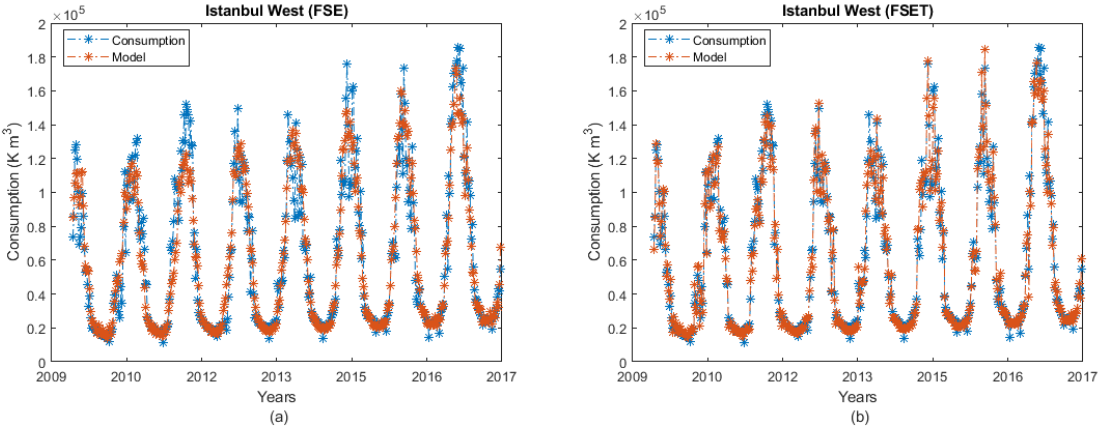

**Figure 5:** Weekly forecasting for Istanbul West using (a) FSE and (b) FSET

### 3.3 Modeling of Daily Variations

Annual and monthly consumption patterns are needed for long term take-or-pay contracts. On the other hand, the liberalization of natural gas market necessitates short-term predictions on a daily basis. In this section, we study the modelling and forecast of daily consumption by using a linear regression supplemented by an error correction scheme. Figure 6 (a) and (b) present daily estimation profiles for FSE and FSET, respectively.

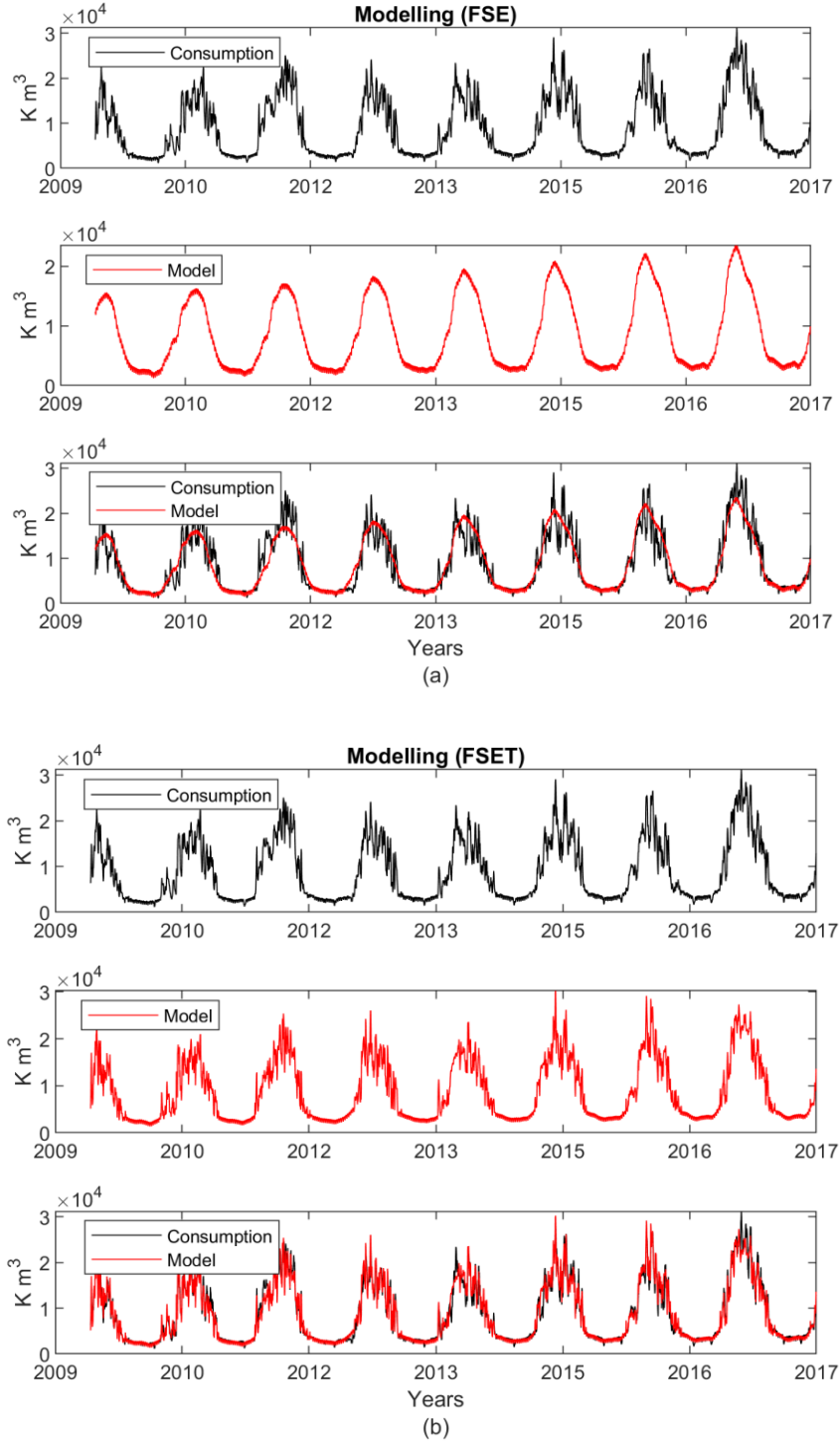

**Figure 6:** Forecasting for Istanbul West using (a) FSE, (b) FSET

The daily demand it is expected to be more variable compared to weekly and monthly demand profiles. In all cases, the FSE model error rates are unsatisfactory, as it can be seen from an inspection of the corresponding graphs. Table 1 summarizes the performance of the forecasting model FSET for each time scale.

The relatively high error rates are due to the strong dependency of the natural gas consumption on weather conditions. In fact, a model that consists of Fourier series only can represent quite faithfully time series data whose main components depend on solar cycles. For example, in Yukseltan et al. (2015) the aggregate electricity consumption in Turkey is modeled by a modulated expansion in Fourier series with modeling errors as low as 3-4%, without the need to use any external information such as temperature. This is basically due to the fact that in Turkey, electricity is used for illumination and cooling, but not widely used for heating. Thus, variations related to the solar cycle are successfully modeled by an FSE model. On the other hand, these types of models would be limited in the modeling of electricity consumption in countries that use electricity for heating. In the case of natural gas, heating requirements dominate household consumption; thus, the FSE model is expected to have limited success and the FSET model needs to be used. Although it would be possible to develop sophisticated models that take into account weather forecasts, their use in the long-term forecast would be limited.

| City | Period | *MAPE* (%) | | | *RMSE* (%) | | |
|---|---|---|---|---|---|---|---|
| | | **Monthly** | **Weekly** | **Daily** | **Monthly** | **Weekly** | **Daily** |
| Istanbul W | 2010-2017 | 5.62 | 8.12 | 10.31 | 5.96 | 8.75 | 12.14 |
| Istanbul E | 2010-2017 | 6.39 | 8.62 | 11.17 | 6.78 | 9.50 | 13.01 |
| Ankara | 2010-2017 | 7.07 | 10.74 | 16.52 | 4.36 | 7.72 | 12.16 |
| Eskisehir | 2010-2015 | 6.82 | 10.11 | 17.05 | 4.35 | 7.97 | 17.40 |

**Table 1:** FSET modelling errors of four major cities for monthly, weekly, and daily forecasting

We see that modeling errors get larger moving from monthly to daily models. Although the error rates are promising and can help plan the natural gas demand, it is possible for prediction errors to be much larger. Thus, we should use alternatives to improve forecasting accuracy.

4. **Forecasting Using Fourier Series Expansion with Temperature and Feedback**

In the previous section, we have shown that errors in modeling daily variations are about 10%, despite the use of temperature data. Prediction errors that are found using only the linear model can be decreased if the model is fed with more updated data as regressor. Therefore, we adopt the approach of supplementing the prediction by a day-ahead forecast using feedback. The accuracy of the forecast is important for deregulated energy markets in which natural gas can be purchased from the market through auctions. If the natural gas demand can be forecast for the days ahead, a proper demand analysis can be made and the supply process can be smooth possibly at a reasonably lower cost.

As high or low temperature times last usually a few days, a combination of linear regression with an auto-regressive model is very useful in such cases. The feedback mechanism, which consists of updating the prediction for the next day by the error of the present day, is essentially equivalent to the simplest form of an auto-regressive model, i.e., AR(1). We develop a feedback method called Fourier Series Expansion with Temperature and Feedback (FSETF) to decrease the forecast error in which we add the demand data of the previous day to the model, and thus make the forecasting more realistic. The matrix $F$ given in Eq. (2) is updated with the inclusion of demand of the previous day, and $a$, $y$, $a_1$, and $y_1$ are updated consecutively. The day-ahead feedback increases the demand accuracy and helps FSETF capture the uncertainties which cannot be determined otherwise, as shown in Figure 7.

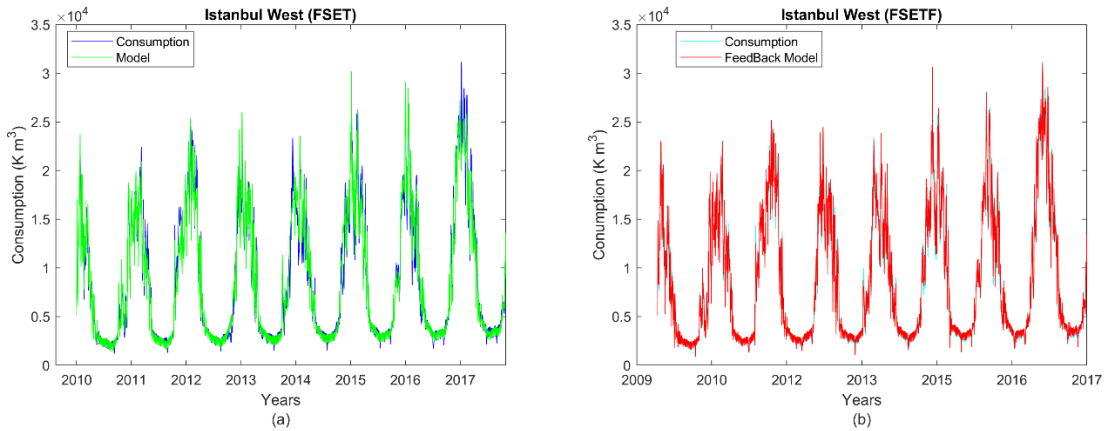

**Figure 7:** Forecasting for Istanbul West using (a) FSET and (b) FSETF

In Table 2, we present the error rates for Istanbul West of FSET and FSETF. Clearly, the feedback mechanism significantly increases the accuracy and decreases the modeling error. Table 3 shows RMSE and MAPE values of the annual forecast for Istanbul West for the years 2014-2016 after the feedback mechanism is applied, and they are quite satisfactory. Note that the error rates are for daily demand forecasting; and for weekly, monthly, and annual forecasting better results should be expected. This scheme is representative of the case where there is no storage capacity and consumption has to be predicted on a daily basis, as in the case of forecasting the demand for electricity.

| City | Period | Model | $MAPE$ (%) | $RMSE$ (%) |
|---|---|---|---|---|
| Istanbul W | 2010-2017 | FSET | 10.3092 | 12.1369 |
| Istanbul W | 2010-2017 | FSETF | 6.1429 | 7.1852 |

**Table 2:** Modeling error of daily forecasting for Istanbul West using FSET and FSETF

| City | Year | *MAPE* (%) | *RMSE* (%) |
|---|---|---|---|
| Istanbul W | 2014 | 6.35 | 9.2979 |
| Istanbul W | 2015 | 5.7252 | 7.0224 |
| Istanbul W | 2016 | 6.5267 | 7.8619 |

**Table 3:** Roll-over forecasting errors of daily data for Istanbul West using FSETF

In order to compare results of FSETF with autoregressive modelling (AR), the partial auto correlation graph with different lags is plotted. The AR(3) model is chosen as a benchmark and the results are shown in Table 4 for Istanbul West. In order to improve AR modelling results, the AR coefficient is calculated again at the end of every year for the next year, so, the model runs with the most recent correlation coefficients. The changes in correlation coefficients are given in Table 5 with respect to years.

| City | Year | FSETF | | AR | |
|---|---|---|---|---|---|
| | | *MAPE* (%) | *RMSE* (%) | *MAPE* (%) | *RMSE* (%) |
| Istanbul W | 2014 | 6.35 | 9.2979 | 9.6605 | 12.9677 |
| Istanbul W | 2015 | 5.7252 | 7.0224 | 8.5298 | 10.7265 |
| Istanbul W | 2016 | 6.5267 | 7.8619 | 8.2068 | 10.0546 |

**Table 4:** Roll-over daily forecasting errors and AR(3) models results.

| City | Year | Lag 1 | Lag 2 | Lag 3 |
|---|---|---|---|---|
| Istanbul W | 2014 | 1.3244 | -0.5227 | 0.1917 |
| Istanbul W | 2015 | 1.0213 | -0.1568 | 0.1323 |
| Istanbul W | 2016 | 1.2057 | -0.2019 | -0.0095 |

**Table 5:** AR Lag coefficients for each year

Even though AR modelling is chosen as the benchmark due to its popularity in the literature, it is not as functional as the FSETF model. FSETF model can be used to forecast different periods; short, medium, and long. However, the AR method only allows forecasting for the short term. Our aim is to draw a projection to suppliers and customers so medium- or long-term projection can be used for various applications and provide more insights to shape future operations.

### 5. Demand Segregation

The consumption profile for summer months that displays weekend effects might allow us to analyze the share of the industrial and residential demand, as heating related residential use is at its minimum during these months. Furthermore, it is known that, in Turkey, the two Muslim religious holidays are the times during which most of the industrial plants are shut down. The timing of these holidays is determined according to the lunar calendar, and they shift by 10 days each year. From 2006 onwards, these holidays coincide with summer months; hence, the holiday effect is visible only after that period. In previous work (Yukseltan et al., 2015), we had used hourly electricity demand for the whole country to show that the ratio of industrial to residential use of electricity was 0.49 in the years 2012-2014. Here, we make a similar analysis using natural gas data. We divide the total consumption into three categories: $R$ denotes pure residential consumption, and $I$ denotes the total industrial consumption. Let $WD$, $WE$ and $H$ be the consumption on weekdays, weekends and holidays, respectively. Let's denote the residential consumption by $R$ an industrial consumption by $I$. During religious holidays, practically all industrial plants are shut down, while during weekends, there is a portion that we denote as $I_0$ continuing to operate. Thus, the weekday consumption $WD$ is $WD=R+I$, the weekend consumption $WE$ is $WE=R+I_0$ and the holiday consumption $H$ is $H=R$. Then, the ratio of weekday to holiday consumption is given as follows.

$$\frac{WD}{H} = \frac{R+I}{R} = 1 + \left(\frac{I}{R}\right) \tag{13}$$

As a second measure, the ratio of weekday to weekend consumption is given below.

$$\frac{WD}{WE} = \frac{R+I}{R+I_0} = \frac{R+I_0+(I-I_0)}{R+I_0} = 1 + \left(\frac{I-I_0}{R+I_0}\right) \tag{14}$$

In Table 4 below, we list the proportion of industrial consumption, $I/R$, for Istanbul West, Istanbul East, Ankara, Bursa, and Eskisehir, for the years 2011-2017, according to the aforementioned two ratios. The significant increase of the proportion of industrial consumption in Bursa in 2016 and 2017 is especially noteworthy.

| | Year | Istanbul W | Istanbul E | Ankara | Bursa | Eskisehir |
|---|---|---|---|---|---|---|
| $\frac{WD}{R} - 1$ (%) | 2011 | 81.19 | 66.52 | 71.40 | 19.75 | 90.07 |
| | 2012 | 75.99 | 66.07 | 66.37 | 65.83 | 17.21 |
| | 2013 | 82.26 | 89.29 | 77.83 | 73.05 | 39.53 |
| | 2014 | 87.67 | 89.08 | 92.07 | 104.84 | 49.59 |
| | 2015 | 71.34 | 62.32 | 56.17 | 46.58 | 18.21 |
| | 2016 | 84.70 | 82.38 | 88.02 | **316.41** | 87.97 |
| | 2017 | 50.81 | 41.46 | 45.90 | **216.90** | 43.42 |
| $\frac{WD}{WE} - 1$ (%) | Year | Istanbul W | Istanbul E | Ankara | Bursa | Eskisehir |
| | 2011 | 21.15 | 14.69 | 7.59 | 12.07 | 2.11 |
| | 2012 | 23.15 | 15.88 | 12.11 | 14.43 | 10.36 |
| | 2013 | 20.23 | 14.02 | 12.27 | 20.87 | 8.16 |

| | 2014 | 21.81 | 14.96 | 12.07 | 32.02 | 12.22 |
| | 2015 | 22.84 | 15.47 | 11.71 | 7.34 | 7.62 |
| | 2016 | 18.84 | 12.57 | 8.10 | **60.92** | 10.62 |
| | 2017 | 13.81 | 7.37 | 7.18 | **44.79** | 10.98 |

**Table 6:** The proportion of industrial natural gas consumption according to two measures in five cities in years 2011-2017

We analyzed the average demand for weekdays and weekends in both summers and winters for the past data. Industrial use is at its lowest level on religious and public holidays, and a comparative approach showed us that one can use the proposed methodology to roughly estimate the ratio of industrial and residential demand in each city. This information will also help to plan demand if a change in industrial production is expected as a result of economic impacts. The effect of industrial demand on total demand is obvious and dominant in some well-industrialized cities. Figure 8 shows the demand for Bursa and Eskisehir for the past years. We would like to note that in the summer of 2017, the religious holiday spanned the period of 24-27 June. The weekend consumption for the following two weeks in Bursa surpassed the weekday levels, possibly to compensate for the labor loss during the holiday. As this is an exceptional case, we excluded this data from our analysis. Nevertheless, the increasing industrial activities in the last two years still caused a high jump in total demand, as observed in this figure. This shows that industrial demand is more dominant in determining the total demand for these cities.

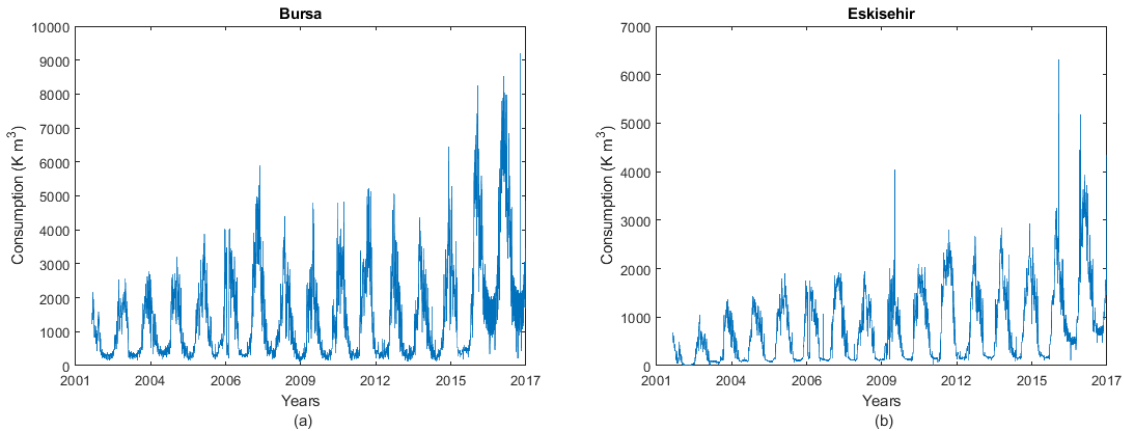

**Figure 8**: The consumption for (a) Bursa and (b) Eskisehir

### 6. Conclusions

Natural gas is an important commodity that is crucial for economic growth and daily life. It is imported from energy-rich countries through long-term take-or-pay contracts. Many companies pay more than they should have paid for the natural gas they did not consume as a result of the take-or-pay conditions. On the other hand, it is also possible to experience shortages if more natural gas is consumed than the ordered amount. A proper planning methodology and consumption forecasting minimize the risk of shortage or extra payment. In this paper, we propose FSE and FSET models, as an expansion in Fourier series; and the temperature is added as a regressor in the latter. We have shown that temperature has a significant role in determining the demand. Although we have generated quite a satisfactory forecasting profile with these methods, we have added a day-ahead feedback mechanism to increase the forecasting accuracy and proposed the FSETF model. The MAPE and RMSE values are at acceptable levels comparing to time series model, and we have shown that there is a gradual improvement moving from FSE to FSET and to FSETF. The models are able to generate long-term demand forecasts, which are crucial for the planning of supply. They are also able to forecast daily, weekly, and monthly consumption that are important parameters for a decision maker, as the consumers need to submit their daily, weekly, and monthly natural gas demand in the contract.

The number of players in the system is expected to increase, and day-ahead and intra-day markets will start functioning in the near feature. In a fully deregulated natural gas market, such forecasting models can be more important as daily buy-sell decisions require an accurate demand forecasting methodology. This model will help make predictions in different time periods in the long and short-terms by considering predetermined parameters. Every consumer can add their own parameter and see its effect on consumption to minimize their deviation from the submitted demand to have a smooth operation. It is also worth mentioning that industrial demand might dominate residential demand for some cities. We have shown that it is possible to analyze the role of industrial demand to determine the total demand. The national holidays in which the industrial production is minimized and summer days in which residential consumption is limited provide valuable information for determining the share of industrial demand in total demand. This research presents a useful policy recommendation for decision-makers in terms of benefiting from the past demand data, temperature information, and industrial production.

Modelling natural gas consumption and making long-term predictions also give an idea about the quantity that can be possibly consumed. The models enable decision makers to observe and add several parameters that affect natural gas consumption. Also, using Fourier expansion helps to catch periodic patterns of consumption. To improve the model in future, it is possible to include additional parameters such as the number of consumers in the system, the information for the weather forecast and the availability of storage for natural gas as well.

**Funding**: This research did not receive any specific grant from funding agencies in the public, commercial, or not-for-profit sectors.

**CRediT author statement:**
**Ergun Yukseltan**: Methodology, Validation, Analysis, Investigation, Formal Analysis
**Ahmet Yucekaya**: Conceptualization, Validation, Analysis, Supervision, Administration
**Ayse Bilge**: Methodology, Validation, Analysis, Writing - Review & Editing, Formal Analysis
**Esra Agca**: Methodology, Validation, Analysis, Writing - Review & Editing